\documentclass[acus]{JAC2003}

\usepackage{graphicx}
\usepackage{amssymb}
\usepackage{amsmath}
\usepackage{bm}
\usepackage{url}
\usepackage{epstopdf}

\begin{document}

\title{INVESTIGATION OF A 2-COLOUR UNDULATOR FEL USING PUFFIN}
\author{L.T. Campbell$^{1,2}$, B.W.J. M$^{\mathrm c}$Neil$^1$ and S. Reiche$^3$ \\ 
$^1$SUPA, Department of Physics, University of Strathclyde, Glasgow, UK, \\ 
$^2$ASTeC, STFC Daresbury Laboratory and Cockcroft Institute, Warrington WA4 4AD, United Kingdom\\
$^3$Paul Scherrer Institute, 5232 Villigen PSI, Switzerland}

\date{\today}

\maketitle

\begin{abstract}
Initial studies of a 2-colour FEL amplifier using one monoenergetic electron beam are presented. The interaction is modelled using the unaveraged, broadband FEL code Puffin. A series of undulator modules are tuned to generate two resonant frequencies along the FEL interaction and a self-consistent 2-colour FEL interaction at widely spaced non-harmonic wavelengths at 1nm and 2.4nm is demonstrated. 
\end{abstract}

\section{Introduction}

With X-ray SASE FEL's now in operation around the world, there is now user interest  in the simultaneous delivery of two distinct wavelengths. This may be possible by preparing the beam before injection into the undulator, or by using two electron beams of different energy, known as a two-beam or two-stream FEL. Another method, the so-called 2-colour FEL,  generates two radiation wavelengths simultaneously from a single mono-energetic electron beam by injecting it through alternative undulator modules that are tuned to the two different wavelengths. There may also be a need for more broadband broadband sources, which may be achievable using an energy chirped electron beam in a 2-colour FEL configuration, and matching together the spectra generated by each of the undulator modules. 

Averaged FEL simulation codes such as~\cite{MEDUSSA,GENESIS,FAST,GINGER} that make the Slowly Varying Envelope Approximation (SVEA)~\cite{SVEA}, can readily model 2-colour FEL interactions when the radiation wavelengths are about a fundamental wavelength $\lambda_1$ and e.g. its third harmonic $\lambda_3 = \lambda_1/3$. This requires the radiation to be described by two distinct computational fields as the field sampling at any resonant radiation wavelength $\lambda_r$ means the frequency range able to be sampled is limited by the Nyquist condition to $\omega_r/2<\omega<3\omega_r/2$. Furthermore, the accuracy of the model decreases the further away from the central resonant frequency $\omega_r$.   For a fundamental and say its 3rd harmonic field, the sampling can be over a common set of `slices' each an integer number of fundamental wavelengths in length. 
However, for cases where the two wavelengths in a 2-colour FEL  are strongly non-harmonic and well separated in frequency, e.g. where $\lambda_2 = 2.4 \times \lambda_1$,  modeling of two computational fields using the averaged SVEA approximation becomes problematic -  integer wavelength long slices of the two fields cannot be coincident and so when radiation propagation effects are modeled, electrons within these different length slices are driven asymmetrically in phase space leading to unphysical instabilities.

The unaveraged free electron laser simulation code Puffin~\cite{POP} does not make the SVEA approximation and is therefore not restricted to the limited bandwidth of conventional averaged FEL codes. Furthermore, because Puffin  also models the electron beam dynamics without  averaging,  the electrons are not confined to localised `slices' of width the fundamental  radiation wavelength, and  a realistic electron beam interaction can occur over a broad bandwidth of radiation wavelengths. Puffin (along with other unaveraged codes e.g.~\cite{muffin,femfel,hooker,maroli}) is then ideally suited to simulating and investigating the physics of a multi-color FELs.

In what follows, the Puffin code was  modified to allow for undulator modules of different undulator parameters $\bar{a}_w$. This code was then used to simulate a high-gain 2-colour FEL amplifier at distinct, well separated, non-harmonic wavelengths. It is not immediately clear how such an interaction would be expected progress and in particular how the electron phase space in interacting resonantly with two distinct radiation wavelengths would develop. The interaction is therefore first seeded, rather than starting from noise, to allow a cleaner picture of the electron beam and radiation evolution in the 2-colour interaction. An example is then given of an unseeded, 2-colour SASE FEL interaction.  

\section{Mathematical Model}

Puffin uses a system of equations which utilizes scaled variables developed in~\cite{bnp}. These dimensionless variables are scaled with respect to the FEL parameter, defined as
\begin{align}
\label{rho}
\rho=\frac{1}{\gamma_r}\left(\frac{\bar{a}_w \omega_p}{4ck_w}\right)^{2/3},
\end{align}
where $\bar{a}_w \propto B_w\lambda_w$ is the usual rms undulator parameter, $B_w$ is the rms undulator magnetic field strength, $\lambda_w$ is the undulator period and $k_w = 2\pi / \lambda_w$. In the cases simulated in this paper, the undulator tuning of $\bar{a}_w$ is obtained by varying the rms undulator magnetic field alone. This tuning therefore also changes the  FEL parameter  and it is convenient to re-scale the equations in a general way for the different undulator module tunings by introducing  the undulator module dependent parameter   $\alpha \equiv \bar{a}_w /  \bar{a}_{w1}$, where the sub-script `1' refers to the value of $\bar{a}_w$ of the first module. By similarly re-defining  
\begin{align}
\rho = \alpha^{2/3} \rho_1, \label{rhon}
\end{align}
again where sub-script `1' refers to the values of the first module, the parameter $\alpha$ becomes an explicit parameter in the equations. A constant focussing channel, with strength expressed as a fraction of the natural undulator focussing which is a function of $\bar{a}_w$~\cite{POP}, maintains a constant electron beam radius for different modules by varying the focussing factor so that $f=\alpha f_1$. The net effect of these scalings maintains the exact form of the scaled equations of~\cite{POP} with the exception of two multiplicative factors of $\alpha$ in the equations describing transverse and longitudinal motion of the electrons. 
The system of equations obtained is  general, and $\alpha$ can be made any function of distance through the interaction as required. It could, for example, be used to taper the undulator in Puffin, or to provide any number of differently tuned undulator modules. For the purposes of this paper, however, $\alpha$ is varied between only two values corresponding to the two colours of radiation required. Further details of the scaling will be described elsewhere.

It is also noted here that in the 2-colour FEL, with only one resonant wavelength interaction occurring in each undulator module, radiation at the non-resonant wavelength will effectively propagate in free space. Two effects can be expected from this. Firstly, the non-resonant wavelength in an undulator module  will undergo free-space diffraction (unless it is an harmonic when some guiding can occur~\cite{psase}), which will reduce the net coupling of the radiation to the electron beam and so  increase the effective gain length of the interaction for both colours. Secondly, the addition of extra relative slippage between the radiation and the electrons in the non-resonant undulator modules may affect the spectrum by introducing modal effects~\cite{mlsase} and may also reduce the radiation bandwidth~\cite{ipac,psase,hbsase}.

\section{Simulation Results}
Simulation results using this modified version of Puffin are now presented that model a 2-colour FEL interaction using a single electron beam in sequential undulators tuned to two different wavelengths. The first example is seeded from some temporally coherent external laser source at both frequencies - this results in a simpler (cleaner)  electron beam phase space. The second example starts up from the electron beam shot-noise i.e. 2-color SASE. Radiation output spectra for differently tuned 2-colour FELs are then summarised. 

In a 2-color FEL, there will  be a small drift section between each undulator module that may require radiation/electron beam phase matching using small tuning chicanes. Although Puffin is capable of modelling the drift sections none are modelled here and an instantaneous change in undulator parameter is applied. Puffin is also used here in the 1D limit as described in \cite{POP}, so that simulations cost considerably less computational effort and the 2-colour interaction can be observed in its simplest form. More detailed effects including diffraction, that can be expected to alter the optimum set-up, and will be investigated in future work.

\subsection{Seeded Example}
Equation (\ref{rhon}) defines how $\rho$ changes with different undulator tunings. At longer wavelengths (larger $\bar{a}_w$) a shorter gain length (at least in the 1D limit) results in saturation in a shorter interaction length. The simplest FEL configuration will have undulator modules with a fixed number of periods, limiting somewhat the options for driving both wavelengths to the same intensity. 
\begin{figure}
\centering
\includegraphics[width=85mm]{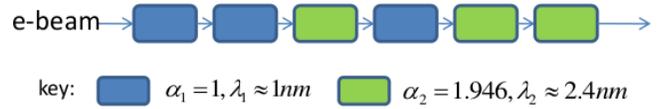}
\caption[Undulator layout]{Modular undulator layout for 2-colour operation.}
\label{figure1}
\end{figure}
To compensate here for the difference in gain length of the two colours, the $N_m=6$ undulator modules are alternated in the configuration of Fig.\ \ref{figure1}. This allows both wavelengths to receive gain as the interaction progresses unlike other options that amplify each wavelength in two consecutive interactions. The method used here stops the longer wavelength radiation dominating the interaction at the initial stages by inducing electron energy spread that  inhibits amplification of the shorter wavelength, longer gain length  interaction. It can be expected from the larger $\rho$ value for the longer wavelength, and the equal total interaction length for both wavelengths, that the longer wavelength will have the larger saturation intensity~\cite{bnp}. It is preferable, therefore, that the shorter wavelength interaction, which induces a smaller electron energy spread than that at the longer wavelength, saturates first, thereby allowing the longer wavelength to extract more energy in the final undulator modules. For this reason it is preferable to end the interaction using undulator modules resonant at the longer wavelength . 
\begin{table}
\centering
\caption[Table of parameters]{Table of parameters for 2-colour simulations}
\begin{tabular}{|  c   | c  |   c   |  c  |}
\hline
& Seeded & SASE & SASE scan \\
\hline
$\rho$ & $0.005$ & $0.005$ & $0.001$ \\
\hline
$\gamma$ & $6300$ & $6300$ & $6900$ \\
\hline
$\sigma_\gamma/\gamma$ & $10^{-4}$ & $10^{-4}$ & $10^{-4}$ \\
\hline
$\bar{k}_{\beta}$ & $1.12 \times 10^{-2}$ & $1.12 \times 10^{-2}$ & $1.12 \times 10^{-2}$ \\
\hline
$N_m$ & $6$ & $13$ & $13$ \\ 
\hline
$N_w$ & $25$ & $25$ & $100$ \\ 
\hline
$\alpha_1$ & $1$ & $1$ & $1$ \\ 
\hline
$\alpha_2$ & $1.946$ & $1.946$ & scan \\
\hline
\end{tabular}
\label{table}
\end{table}



\begin{figure}
\centering
\includegraphics[width=85mm]{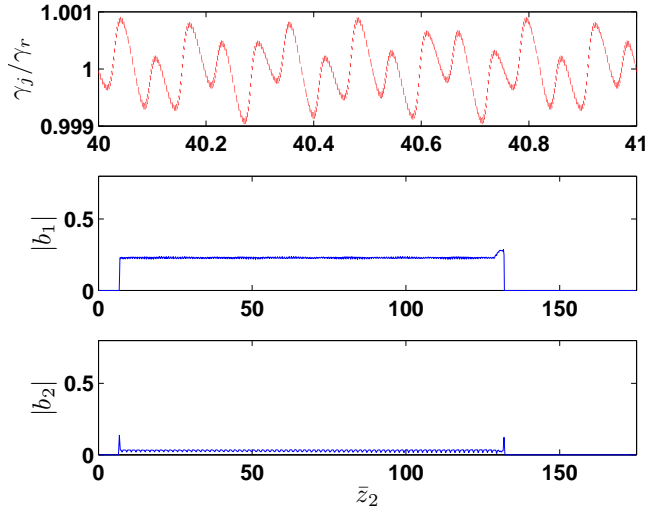}
\caption[Bunching at end of sim]{Electron parameters at output of 3rd undulator module: Detail of electron phase space (top); modulus of bunching parameter at $\lambda_1$ (middle); modulus of bunching parameter at $\lambda_2$ (bottom).}
\label{figure2}
\end{figure}
\begin{figure}
\centering
\includegraphics[width=85mm]{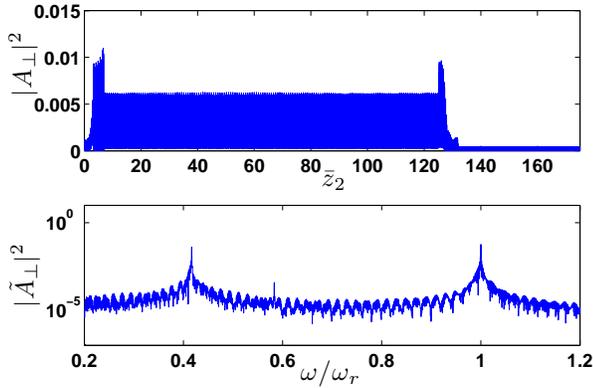}
\caption[Bunching at end of sim]{Scaled radiation intensity (top) and spectrum (bottom) at output of 3rd undulator module.}
\label{figure3}
\end{figure}

The parameters for the first seeded case are shown in Table \ref{table}. A small seed field is injected at each resonant frequency, and the electron beam has a `flat-top' current profile. In the scaled $\bar{z}_2$ frame, the resonant wavenumbers are $\bar{k}_1 = 100$ and $\bar{k}_2 \approx 41.7$.
A bunching parameter for each wavelength can be defined as:
\begin{align}
b_{1,2} = \frac{1}{L_{1,2}} \sum_{k=1}^{N_k} \bar{\chi}_k e^{i\bar{k}_{1,2}\bar{z}_{2k}}
\end{align}
where $N_k$ is the number of macroparticles, and $\bar{\chi}_k$ is the charge weighting factor~\cite{POP}. For each   wavelength, the beam is discretised into slices an integer multiple of resonant wavelengths long,  $L_1$ and $L_2$ respectively, and the bunching parameters $b_{1,2}$ calculated.
Detail of the electron phase space and the modulus of the bunching parameters across the whole beam are shown in Fig.\ \ref{figure2} at the end of the third module. The  scaled temporal and spectral radiation intensity is plotted in Fig.\ \ref{figure3}. 
The head of the electron pulse starts the beginning of the interaction at $\bar{z}_2=0$ in the radiation window, and slips behind the radiation towards larger values of $\bar{z}_2$ as the interaction progresses through the undulator. At the end of the third module, the interaction is coming towards the end of the linear regime, with the interaction (and radiation intensity) at the shorter wavelength being slightly greater, having passed through two resonant undulators. The longer wavelength interaction in the third module is seen to have modified the electron phase space away from the familiar single wavelength FEL interaction, with two distinct energy modulations at the two wavelengths being visible. The effects of Coherent Spontaneous Emission (CSE) in the small enhanced bunching at the tail of the electron bunch at $\bar{z}_2\approx 130$ is also visible~\cite{femfel}.

\begin{figure}[b]
\centering
\includegraphics[width=85mm]{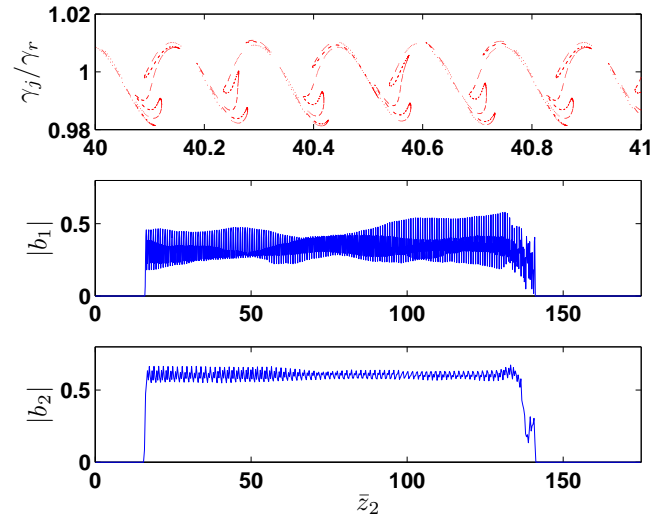}
\caption[Bunching at end of sim]{As Fig.\ \ref{figure2} at output from 6th undulator module.}
\label{figure4}
\end{figure}

\begin{figure}
\centering
\includegraphics[width=85mm]{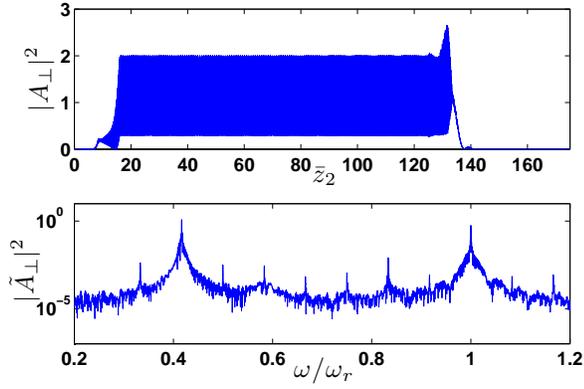}
\caption[Bunching at end of sim]{As Fig.\ \ref{figure3} at output from 6th undulator module.}
\label{figure5}
\end{figure}

Figs.\ \ref{figure4} and \ref{figure5} plot the same at the output of the end of the final undulator module where the longer wavelength is close to saturation, slightly ahead of the shorter.
A quite complex electron phase space had developed as a result of the 2-colour interaction which has generated spectral intensities that are approximately equal for the two colours. This complexity of the phase space bunching results, not just during the resonant interactions with the radiation, but continues due to dispersion of energy modulations during propagation through non-resonant undulator modules. This can work to reduce bunching where the energy modulation has shifted the beam off-resonance. Interestingly, this effect can cause modulations of the bunching parameter in $\bar{z}_2$ at both wavelengths. A weak modal-like structure is also visible in the radiation spectrum. The origin of this is not yet established, but may be due to the modal effects of~\cite{mlsase}.  

\subsection{2-Color SASE Example}

\begin{figure}[b]
\centering
\includegraphics[width=85mm]{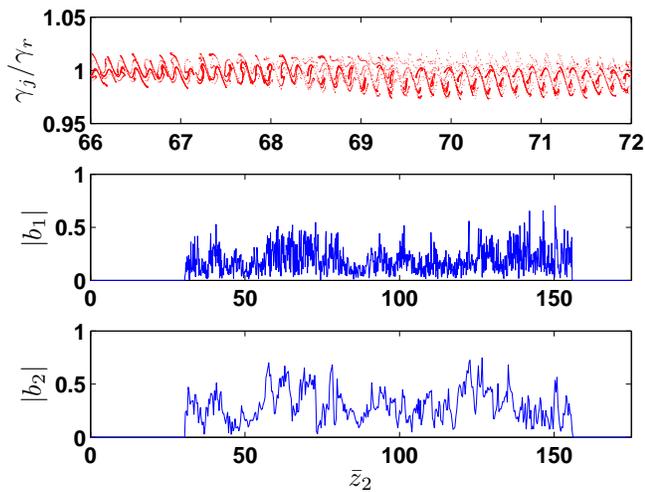}
\caption[Bunching at end of sim]{Electron parameters at output of 13th undulator module in the SASE case: Detail of electron phase space (top); modulus of bunching parameter at $\lambda_1$ (middle); modulus of bunching parameter at $\lambda_2$ (bottom).}
\label{figure6}
\end{figure}

The 2-colour FEL interaction is now modelled starting from noise via SASE.  A similar, but extended, undulator beamline as Fig. \ref{figure1} is used with a $1,1,2,1,1,2,...,1,1,2,2$, undulator module configuration, to allow for the build-up from noise. Otherwise all parameters are identical to the seeded case. Note the extra final undulator module resonant at the longer wavelength, as discussed above,  that allows extraction of further longer wavelength energy from the beam that is saturated to further shorter wavelength emission. Plots similar to the seeded case of above are shown at the output of the $N_m=13$ module beamline in figure \ref{figure6}. 
As the system now starts from noise the resultant beam phase space is more complicated but maintains similar characteristics to the seeded case. The system is close to saturation and the beam is seen to have bunching components  at both wavelengths simultaneously. 

One factor that is present in the 2-colour SASE, that does not exist in the seeded case, is the potential for phase mis-matching between the radiation and the electron beam due to the extra induced slippage in the alternate non-resonant undulator modules. This may be mitigated by using modules that are less than one gain length ensuring relative slippages are less than a cooperation length, so maintaining beam/radiation phase matching.  
\begin{figure}
\centering
\includegraphics[width=85mm]{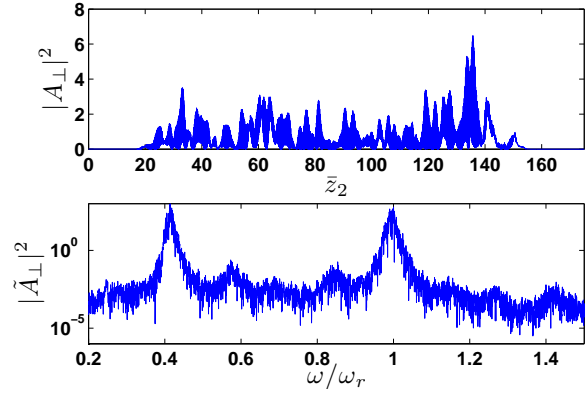}
\caption[Bunching at end of sim]{Scaled radiation intensity (top) and spectrum (bottom) at output of 13th undulator module in the SASE case.}
\label{figure7}
\end{figure}

\begin{figure}
\centering
\includegraphics[width=85mm]{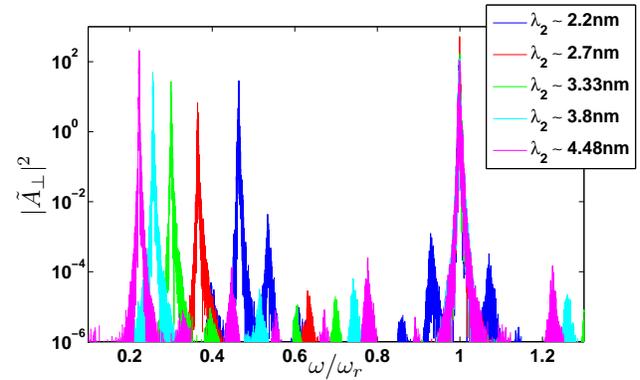}
\caption[Bunching at end of sim]{Scaled radiation intensity spectra at output for a range of 2-colour operating frequencies}
\label{figure8}
\end{figure}

Similar 2-colour SASE simulations were also carried out for different wavelengths. This was achieved by using the extended configuration of Fig.\ \ref{figure1} above, but for different values of $\alpha_2$. The FEL $\rho$ parameter has been reduced to $\rho = 0.001$ for this scan, and the number of undulator periods per module has been increased to $N_w = 100$ to allow for the longer gain length. The electron beam has similar parameters to a shorter wavelength FEL such as that on the Athos FEL proposed at SwissFEL~\cite{swissfel}. The spectra of the scaled output intensities are plotted in Fig. \ref{figure8} for a range of non-harmonic values of $\lambda_2$. 

It is seen that the principle of operation of the 2-colour FEL applies to across a wide range of wavelengths. No attempt was made to optimise the output, so the output at a given wavelength can be expected to be improved upon. The statistical nature of the output is also unknown from these single-shot simulations.

\section{Conclusion}
The principle of a 2-colour FEL amplifier interaction using a single electron beam in an alternating series of differently tuned undulator modules was simulated for widely spaced, non-harmonic wavelengths. This modeling is only practical using non-averaged, broad bandwidth FEL simulation codes such as Puffin. 
Puffin was modified from that described in~\cite{POP}, to allow variation of the undulator magnetic field strength to change the resonant FEL wavelength.  This allows Puffin to model not just a 2-color FEL, but any general variation in undulator field strength across a wide range of resonant wavelengths as the interaction proceeds.  

Both seeded and SASE 2-colour operation were demonstrated to saturation and gave similar peak intensity spectra at both the resonant wavelengths. This simultaneous FEL lasing to saturation by one electron beam at two distinct, harmonically uncoupled, wavelengths is perhaps not an immediately intuitive result. However, the electron phase-space evolution has a rich structure that clearly demonstrates simultaneous electron bunching and emission at the two distinct wavelengths.

This result may open up other avenues for further research, for example in the generation of higher-modal radiation emission, or perhaps frequency mixing process from the FEL. 

In the case of the 2-color SASE simulations, only a limited small of simulations were performed for each case so that the statistical nature of the 2-colour processes were not explored. 

Other factors must be taken into account in future 3D simulations. Diffraction will clearly have an effect during `free' propagation of the radiation in the non-resonant undulator modules. Another factor is the transverse beam matching to different undulator modules. This could become more problematic for larger differences in $\bar{a}_w$ between undulator modules and may limit the range of the 2-colour operation. Clearly, further examination in $3D$ is needed to identify and perhaps find solutions to such issues.

\end{document}